\documentclass[12pt,thmsb]{article}
\usepackage{amsfonts}

\usepackage{amsmath}
\usepackage{harvard}

\advance \topmargin by -\headheight
\advance \topmargin by -\headsep
\evensidemargin \oddsidemargin
\let\tilde=\widetilde

\input tcilatex
\makeatletter
\def\@biblabel#1{\hspace*{-\labelsep}}
\makeatother

\begin{document}

\title{Structural Breaks in Time Series\thanks{%
We wish to thank Palgrave Macmillan for granting permission to use parts of
the following prior review: Perron, P. ``Structural Change''\ in \textit{The
New Palgrave Dictionary of Economics}, 2nd ed, S. Durlauf and L. Blume
(eds.), 2008, Palgrave Macmillan. }}
\author{Alessandro Casini \textbf{\thanks{%
Economics Department, Boston University, 270 Bay State Rd., Boston MA 02215
(acasini@bu.edu).}} \\
Boston University \and Pierre Perron\textbf{\thanks{%
Economics Department, Boston University, 270 Bay State Rd., Boston MA 02215
(perron@bu.edu).}} \\
Boston University}
\date{May 9, 2018}
\maketitle

\begin{abstract}
This chapter covers methodological issues related to estimation, testing and
computation for models involving structural changes. Our aim is to review
developments as they relate to econometric applications based on linear
models. Substantial advances have been made to cover models at a level of
generality that allow a host of interesting practical applications. These
include models with general stationary regressors and errors that can
exhibit temporal dependence and heteroskedasticity, models with trending
variables and possible unit roots and cointegrated models, among others.
Advances have been made pertaining to computational aspects of constructing
estimates, their limit distributions, tests for structural changes, and
methods to determine the number of changes present. A variety of topics are
covered. The first part summarizes and updates developments described in an
earlier review, Perron (2006), with the exposition following heavily that of
Perron (2008). Additions are included for recent developments: testing for
common breaks, models with endogenous regressors (emphasizing that simply
using least-squares is preferable over instrumental variables methods),
quantile regressions, methods based on Lasso, panel data models, testing for
changes in forecast accuracy, factors models and methods of inference based
on a continuous records asymptotic framework. Our focus is on the so-called
off-line methods whereby one wants to retrospectively test for breaks in a
given sample of data and form confidence intervals about the break dates.
The aim is to provide the readers with an overview of methods that are of
direct usefulness in practice as opposed to issues that are mostly of
theoretical interest.\vspace{0.2in}

\textbf{JEL Classification}: C22

\textbf{Keywords}: Change-point, linear models, testing, confidence
intervals, trends, stationary and integrated regressors, factor models,
Lasso, foracasts.
\end{abstract}

\thispagestyle{empty}\setcounter{page}{0}\baselineskip=18.0pt\newpage

\pagestyle{plain}

This chapter covers methodological issues related to estimation, testing and
computation for models involving structural changes. The amount of work on
this subject is truly voluminous and any survey is bound to focus on
specific aspects. Our aim is to review developments as they relate to
econometric applications based on linear models. Substantial advances have
been made to cover models at a level of generality that allow a host of
interesting practical applications. These include models with general
stationary regressors and errors that can exhibit temporal dependence and
heteroskedasticity, models with trending variables and possible unit roots
and cointegrated models, among others. Advances have been made pertaining to
computational aspects of constructing estimates, their limit distributions,
testing and methods to determine the number of changes present. The first
part summarizes and updates developments described in an earlier review,
Perron (2006), with the exposition following heavily that of Perron (2008).
Additions are included for recent developments: testing for common breaks,
models with endogenous regressors (emphasizing that simply using
least-squares is preferable over instrumental variables methods), quantile
regressions, methods based on Lasso, panel data models, testing for changes
in forecast accuracy, factors models and methods of inference based on a
continuous records asymptotic framework. Our focus is solely on linear
models and deals with so-called off-line methods whereby one wants to
retrospectively test for breaks in a given sample of data and form
confidence intervals about the break dates. Given the space constraint, our
review is obviously selective. The aim is to provide an overview of methods
that are of direct usefulness in practice as opposed to issues that are
mostly of theoretical interest.

\noindent \textbf{The basic setup}. We consider the linear regression with $%
m $ breaks (or $m+1$ regimes): 
\begin{equation}
y_{t}=x_{t}^{\prime }\beta +z_{t}^{\prime }\delta _{j}+u_{t},\hspace{0.4in}%
t=T_{j-1}+1,...,T_{j},  \label{model1}
\end{equation}%
for $j=1,...,m+1$, following Bai and Perron (1998) (henceforth BP). In this
model, $y_{t}$ is the observed dependent variable; $x_{t}$ ($p\times 1$) and 
$z_{t}$ ($q\times 1$) are vectors of covariates and $\beta $ and $\delta
_{j}~(j=1,...,m+1)$ are the corresponding vectors of coefficients; $u_{t}$
is the disturbance. The break dates $(T_{1},...,T_{m})$ are unknown (the
convention that $T_{0}=0$ and $T_{m+1}=T$ is used). The aim is to estimate
the regression coefficients and the break points when $T$ observations on $%
(y_{t},x_{t},z_{t})$ are available. This is a partial structural change
model since the parameter vector $\beta $ is not subject to shifts. When $%
p=0 $, we obtain a pure structural change model with all coefficients
subject to change. A partial structural change model can be beneficial in
terms of obtaining more precise estimates and more powerful tests. The
method of estimation is standard least-squares (OLS), i.e., minimizing the
overall sum of squared residuals (SSR) $\sum_{i=1}^{m+1}%
\sum_{t=T_{i-1}+1}^{T_{i}}[y_{t}-x_{t}^{\prime }\beta -z_{t}^{\prime }\delta
_{i}]^{2}$. Let $\hat{\beta}(\{T_{j}\})$ and $\hat{\delta}(\{T_{j}\})$
denote the estimates based a partition $(T_{1},...,T_{m})$ denoted $%
\{T_{j}\} $. Substituting these in the objective function and denoting the
resulting SSR as $S_{T}(T_{1},...,T_{m})$, the estimated break points are%
\begin{equation}
(\hat{T}_{1},...,\hat{T}_{m})=\mathrm{argmin}%
_{(T_{1},...,T_{m})}S_{T}(T_{1},...,T_{m}),  \label{min}
\end{equation}%
with the minimization taken over a set of admissible partitions (see below).
The parameter estimates are those associated with the partition $\{\hat{T}%
_{j}\}$, i.e., $\hat{\beta}=\hat{\beta}(\{\hat{T}_{j}\})$, $\hat{\delta}=%
\hat{\delta}(\{\hat{T}_{j}\})$. Since estimation is based on OLS, even if
changes in the variance of $u_{t}$ are allowed, provided they occur at the
same dates $(T_{1},...,T_{m})$, they are not exploited to increase the
precision of the break date estimators unless a quasi-likelihood framework
is adopted, see below.

\noindent \textbf{The theoretical framework,} \textbf{the assumptions and
their relevance. }To obtain theoretical results about the consistency and
limit distribution of the estimates of the break dates, some conditions need
to be imposed on the asymptotic framework, the regressors, the errors, the
set of admissible partitions and the break dates. By far the most common
asymptotic framework is one whereby as $T$ increases the total span of the
data increases such that the length of the regimes increases
proportionately, which implies that the break dates are asymptotically
distinct, i.e., $T_{i}^{0}=[T\lambda _{i}^{0}]$, where $0<\lambda
_{1}^{0}<...<\lambda _{m}^{0}<1$ (a recent alternative framework proposed is
to let the span fixed and increase the number of observations by letting the
sampling interval decrease; see below). To our knowledge, the most general
set of assumptions in the case of weakly stationary, or mixing, regressors
and errors are those in Perron and Qu (2006). Along with the asymptotic
framework, this implies that what is relevant for inference about a break
date is only the neighborhood around the break date considered. Some
conditions are technical while others restrict the potential applicability
of the results. The assumptions on the regressors specifies that for $%
w_{t}=(x_{t}^{\prime },z_{t}^{\prime })^{\prime }$, $(1/l_{i})%
\sum_{t=T_{i}^{0}+1}^{T_{i}^{0}+[l_{i}v]}w_{t}w_{t}^{\prime }\rightarrow
_{p}Q_{i}(v)$ a non-random positive definite matrix uniformly in $v\in \left[
0,1\right] $. It allows their distribution to vary across regimes but
requires the data to be weakly stationary stochastic processes. This can be
relaxed on a case by case basis though the proofs then depend on the nature
of the relaxation. For instance the scaling used forbids trending
regressors, unless they are of the form $\{1,(t/T),...,(t/T)^{p}\}$, say,
for a polynomial trend of order $p$. Casting trend functions in this form
can deliver useful results in many cases. However, there are instances where
specifying trends in unscaled form, i.e., $\{1,t,...,t^{p}\}$, can deliver
much better results, especially if level and trend slope changes occur
jointly. Results using unscaled trends with $p=1$ are presented in Perron
and Zhu (2005). A comparison of their results with other trend
specifications is presented in Deng and Perron (2006). A generalization with
fractionally integrated errors can be found in Chang and Perron (2016).
Another important restriction is implied by the requirement that the limit
be a fixed, as opposed to stochastic, matrix. This precludes integrated
processes as regressors (i.e., unit roots). In the single break case, this
has been relaxed by Bai et al. (1998) who consider structural changes in
cointegrated relationships in a system of equations. Kejriwal and Perron
(2008)\ provide results for multiple structural changes in a single
cointegrating vector. Consistency still applies but the rate of convergence
and limit distributions are different.

The assumptions on $u_{t}$ and $\left\{ w_{t}u_{t}\right\} $ impose mild
restrictions and permit a wide class of potential correlation and
heterogeneity (including conditional heteroskedasticity) and lagged
dependent variables. It rules out errors with unit roots, which can be of
interest; for example when testing for a change in the deterministic
component of the trend function for an integrated series (Perron and Zhu,
2005). The set of conditions is not the weakest possible. For example,
Lavielle and Moulines (2000) allow the errors to be long memory processes
but consider only the case of multiple changes in the mean. It is also
assumed that the minimization problem is taken over all partitions such that 
$T_{i}-T_{i-1}\geq \epsilon T$ for some $\epsilon >0$.\ \ This is not
restrictive in practice since $\epsilon $ can be small. Under these
conditions, the break fractions $\lambda _{i}^{0}$ are consistently
estimated, i.e., $\hat{\lambda}_{i}\equiv (\hat{T}_{i}/T)\rightarrow
_{p}\lambda _{i}^{0}$ and that the rate of convergence is $T$. The estimates
of the break dates are not consistent themselves. The estimates of the other
parameters have the same distribution as would prevail if the break dates
were known. Kejriwal and Perron (2008) obtain similar results with $I(1)$
regressors for a cointegrated model subject to multiple changes, using the
static regression or a dynamic regression augmented with leads and lags of
the first differences of the $I(1)$ regressors.

\noindent \textbf{Allowing for restrictions on the parameters. }Perron and
Qu (2006) consider a broader framework whereby linear restrictions on the
parameters can be imposed. The class of models considered is $%
y_{t}=z_{t}^{\prime }\delta _{j}+u_{t}$ ($t=T_{j-1}+1,...,T_{j}$) where $%
R\delta =r$, with $R$ a $k$ by $(m+1)q$ matrix with rank $k$ and $r$, a $k$
dimensional vector of constants. The assumptions are the same as discussed
above. There is no need for a distinction between variables whose
coefficients are allowed to change and those whose coefficients are not
allowed to change. Restricting some coefficients to be identical across
regimes can yield a partial structural change model. This is a useful
generalization since it permits a wider class of models of practical
interests; e.g., a model with a specific number of states less than the
number of regimes, or one where a subset of coefficients may be allowed to
change over only a limited number of regimes. They show that the same
consistency and rate of convergence results hold. Moreover, the limit
distributions of the estimates of the break dates are unaffected by the
imposition of valid restrictions, but improvements can be obtained in finite
samples. The main advantages of imposing restrictions are that more powerful
tests and more precise estimates are obtained.

\noindent \textbf{Method to Compute Global Minimizers. }To estimate the
model, we need global minimizers of the objective function (\ref{min}). A
standard grid search requires least squares operations of order $O(T^{m})$,
which is prohibitive when $m>2$. Bai and Perron (2003a), henceforth
BP-2003a, discuss a method based on a dynamic programming algorithm that is
very efficient (see also Hawkins, 1976). Indeed, the additional computing
time needed to estimate more than two break dates is marginal compared to
the time needed to estimate a two break model. Consider the case of a pure
structural change model. The basic idea is that the total number of possible
segments is at most $T(T+1)/2$ and is therefore of order $O(T^{2})$. One
then needs a method to select which combination of segments yields a minimal
value of the objective function. This is achieved efficiently using a
dynamic programming algorithm. For models with restrictions (including a
partial structural change model), an iterative procedure is available, which
in most cases requires very few iterations. Hence, even with large samples,
the computing cost is small. If the sample is very large, various methods
have been proposed that are of order $O(T)$ in computation time. Given that
they are of less importance in economics, such procedures are not reviewed.
Also, in the context of very large data sets, methods using Lasso, discussed
later, appear more promising.

\noindent \textbf{The limit distribution of the estimates of the break
dates. }With the assumptions on the regressors and errors, and given the
asymptotic framework, the limit distributions of the estimates of the break
dates are independent, so that the analysis is the same as for a single
break. This holds because the distance between each break increases at rate $%
T$, and the mixing conditions on the regressors and errors impose a short
memory property so that distant events are independent. The main results for
this case are those of Bai (1997a) for the single break case and the
extension of BP (1998) for multiple breaks. The limit distribution depends
on: a) the magnitude of the change in coefficients (larger changes leading
to higher precision), b) the limit sample moment matrices of the regressors
for the pre and post break segments; c) the so-called `long-run' variance of 
$\{w_{t}u_{t}\}$, which accounts for serial correlation; d) whether the
regressors are trending or not. In all cases, the nuisance parameters can be
consistently estimated and confidence intervals constructed. For a change of
fixed magnitude the limit distribution depends on the finite sample
distribution of the errors. To get rid of this dependence, the asymptotic
framework is modified with the change in parameters getting smaller as $T$
increases, but slowly enough for the estimated break fraction to remain
consistent. The limit distribution obtained in Bai (1997a) and BP (1998)
applies to the case with no trending regressors. With trending regressors, a
similar result is still possible (assuming trends of the form $(t/T)$); see
Bai (1997a) when $z_{t}$ is a polynomial time trend. For an unscaled linear
trend, see Perron and Zhu (2005). Deng and Perron (2006) show that the
shrinking shift asymptotic framework leads to a poor approximation for a
change in a linear trend and that the limit distribution based on a fixed
magnitude of shift is preferable. In a cointegrating regression with $I(1)$
variables, Kejriwal and Perron (2008) show that the estimated break
fractions are asymptotically dependent so that confidence intervals need to
be constructed jointly. If only the intercept and/or the coefficients of the
stationary regressors are allowed to change, the estimates of the break
dates are asymptotically independent.

Besides the original asymptotic arguments used by Bai (1997a) and BP (1998), 
{Elliott and M\"{u}ller (2007) }propose to invert Nyblom's (1989) statistic
to construct confidence sets, while Eo and Morley (2015) generalized
Siegmund's (1988) method thereby inverting the likelihood-ratio statistic of
Qu and Perron (2007); henceforth ILR. The latter methods were mainly
motivated by the fact that the empirical coverage rates of the confidence
intervals obtained from Bai's (1997) method are below the nominal level with
small breaks. The method of Ellliot and M\"{u}ller (2007) delivers the most
accurate coverage rates, though at the expense of increased average lengths
of the confidence sets especially with large breaks. The length can be very
large (e.g., the whole sample) even with very large breaks; e.g., with
serially correlated errors or with lagged dependent variables. Yamamoto
(2016) propose a modification of the long-run variance estimator that
alleviates this problem, though the method does not apply when lagged
dependent variables are present. The ILR-based confidence sets display a
coverage probability often above the nominal level and this results in an
average length larger than with Bai's (1997a) method. See Chang and Perron
(2017) for a review. Kurozumi and Yamamoto (2015) propose confidence sets
obtained by inverting a test that maximizes some weighted average power
function. Overall, the findings suggest a need for a method that provides
over a wide range of empirically relevant models both good coverage
probabilities and reasonable lengths of confidence sets, especially for all
break sizes, whether large or small. See below for a recent alternative
using a continuous time asymptotic framework and the concept of Highest
Density Regions (Casini and Perron, 2017a).

\noindent \textbf{Estimating Breaks one at a time. }Bai (1997b) and BP
(1998) show that one can consistently estimate all break fractions
sequentially. When estimating a single break model in the presence of
multiple breaks, the estimate of the break fraction will converge to one of
the true break fractions, the one that allows the greatest reduction in the
SSR. Then, allowing for a break at the estimated value, a second one break
model can be applied which will consistently estimate the second dominating
break, and so on. Interestingly, Yang (2017) shows that this result fails to
hold for breaks in a linear trend model. Bai (1997b) considers the limit
distribution of the estimates and shows that they are not the same as those
obtained when estimating all break dates simultaneously. Except for the last
break, the limit distributions depend on the parameters in all segments. He
suggests a repartition procedure, which re-estimates each break date
conditional on the adjacent break dates. The limit distribution is then the
same as when the break dates are estimated simultaneously.

\noindent \textbf{Estimation in a system of regressions. }Substantial
efficiency gains can be obtained by casting the analysis in a system of
regressions. Bai et al. (1998) consider estimating a single break date in
multivariate time series allowing stationary or integrated regressors as
well as trends. Bai (2000) considers a segmented stationary VAR\ model with
breaks occuring in the parameters of the conditional mean, the covariance
matrix of the error term or both. The most general framework is that of Qu
and Perron (2007) who consider models of the form $y_{t}=(I\otimes
z_{t}^{\prime })S\beta _{j}+u_{t}$, for $T_{j-1}+1\leq t\leq T_{j}$ $%
(j=1,...,m+1$), where $y_{t}$ is an $n$-vector of dependent variables and $%
z_{t}$ is a $q$-vector that includes the regressors from all equations, and $%
u_{t}\sim (0,\Sigma _{j})$. The matrix $S$ is of dimension $nq$ by $p$ with
full column rank (usually a selection matrix that specifies which regressors
appear in each equation). They also allow for the imposition of a set of $r$
restrictions of the form $g(\beta ,vec(\Sigma ))=0$, where $\beta =(\beta
_{1}^{\prime },...,\beta _{m+1}^{\prime })^{\prime }$, $\Sigma =(\Sigma
_{1},...,\Sigma _{m+1})$ and $g(\cdot )$ is an $r$ dimensional vector. Both
within- and cross-equation restrictions are allowed, and in each case within
or across regimes. The assumptions on the regressors and errors $u_{t}$ are
similar to those discussed above. Hence, the framework permits a wide class
of models including VAR, SUR, linear panel data, change in means of a vector
of stationary processes, etc. Models with integrated regressors are not
permitted. Allowing for restrictions on $\beta _{j}$ and $\Sigma _{j}$
permits a wide range of cases of practical interest: partial structural
change models, block partial structural change models where only a subset of
the equations are subject to change; changes in only some element of the
covariance matrix $\Sigma _{j}$; changes in only the covariance matrix $%
\Sigma _{j}$, while $\beta _{j}$ is the same for all segments; models where
the breaks occur in a particular order across subsets of equations; etc.

The method of estimation is again QML (based on Normal errors) subject to
the restrictions. They derive the consistency, rate of convergence and the
limit distributions of the estimated break dates. Though only root-$T$
consistent estimates of $(\beta ,\Sigma )$ are needed to construct
asymptotically valid confidence intervals, more precise estimates will lead
to better finite sample coverage rates. Hence, it is recommended to use the
estimates obtained imposing the restrictions even though imposing
restrictions does not have a first-order effect on the limiting
distributions of the estimates of the break dates. To make estimation
possible in practice, they present an algorithm which extends the one
discussed in BP (2003a) using an iterative GLS procedure to construct the
likelihood function for all possible segments.

Qu and Perron (2007) also consider ``locally ordered breaks''. This applies
when the breaks across different equations are ``ordered''\ based on prior
knowledge and are ``local''\ since the time span between them is short.
Hence, the breaks cannot be viewed as occurring simultaneously nor as
asymptotically distinct. An estimation algorithm is presented and a
framework to analyze the limit distribution of the estimates is introduced.
Unlike the case with asymptotically distinct breaks, the distributions of
the estimates of the break dates need to be considered jointly. Their
analysis has been considerably extended to cover models with trends and
integrated regresssors in Li and Perron (2017).

\noindent \textbf{Tests that allow for a single break. }To test for a
structural change at an unknown date, Quandt (1960) suggests the likelihood
ratio test evaluated at the break date that maximizes it. This problem was
treated under various degrees of specificity that culminated in the general
treatment by Andrews (1993). The basic method is to use the maximum of the
likelihood ratio test over all possible values of the parameter in some
pre-specified set. For a single change, the statistic is $\sup_{\lambda
_{1}\in \Lambda _{\epsilon }}LR_{T}(\lambda _{1})$, where $LR_{T}(\lambda
_{1})$ denotes the likelihood ratio evaluated at $T_{1}=[T\lambda _{1}]$ and
the maximization is restricted over break fractions in the set $\Lambda
_{\epsilon }=[\epsilon _{1},1-\epsilon _{2}]$ with $\epsilon _{1}$,$\epsilon
_{2}>0$. The limit distribution depends on $\Lambda _{\epsilon }$. Andrews
(1993)\ also considers tests based on the maximal value of the Wald and LM
tests and shows that they are asymptotically equivalent under a sequence of
local alternatives. This does not mean, however, that they all have the same
properties in finite samples. The simulations of Vogelsang (1999), for a
change in mean with serially correlated errors, showed the $\sup LM_{T}$
test to be seriously affected by the problem of non monotonic power, in the
sense that, for a fixed $T$, the power of the test can decrease to zero as
the change in mean increases.

For Model (\ref{model1}) with $i.i.d.$ errors, the LR and Wald tests have
similar properties, so we shall discuss the Wald test. For a single change,
it is defined by (up to a scaling by $q$): 
\begin{equation}
\sup\nolimits_{\lambda _{1}\in \Lambda _{\epsilon }}W_{T}(\lambda
_{1};q)=\sup\nolimits_{\lambda _{1}\in \Lambda _{\epsilon }}\hat{\delta}%
^{\prime }H^{\prime }(H(\bar{Z}^{\prime }M_{X}\bar{Z})^{-1}H^{\prime })^{-1}H%
\hat{\delta}/[SSR_{k}/(T-2q-p)]  \label{ftest}
\end{equation}%
where $\bar{Z}=diag(Z_{1},...,Z_{m+1})$ with $%
Z_{i}=(z_{T_{i-1}+1},...,z_{T_{i}})^{\prime }$, $H$ is such that $(H\delta
)^{\prime }=(\delta _{1}^{\prime }-\delta _{2}^{\prime })$ and $%
M_{X}=I-X(X^{\prime }X)^{-1}X^{\prime }$. Here $SSR_{k}$ is the SSR under
the alternative hypothesis. The break point that maximizes the Wald test is
the same as the estimate obtained by minimizing the SSR provided the
minimization problem (\ref{min}) is restricted to the set $\Lambda
_{\epsilon }$, i.e., $\sup_{\lambda _{1}\in \Lambda _{\epsilon
}}W_{T}(\lambda _{1};q)=W_{T}(\hat{\lambda}_{1};q)$. When serial correlation
and/or heteroskedasticity in the errors is permitted, the Wald test must be
adjusted to account for this, i.e., 
\begin{equation}
W_{T}^{\ast }(\lambda _{1};q)=(T-2q-p)\hat{\delta}^{\prime }H^{\prime }(H%
\hat{V}(\hat{\delta})H^{\prime })^{-1}H\hat{\delta},  \label{robustf}
\end{equation}%
where $\hat{V}(\hat{\delta})$ is robust to serial correlation and
heteroskedasticity; a consistent estimate of 
\begin{equation}
V(\hat{\delta})=\mathrm{plim}_{T\rightarrow \infty }T(\bar{Z}^{\prime }M_{X}%
\bar{Z})^{-1}\bar{Z}^{\prime }M_{X}\Omega M_{X}\bar{Z}(\bar{Z}^{\prime }M_{X}%
\bar{Z})^{-1}.  \label{vhat}
\end{equation}%
Note that it can be constructed allowing identical or different
distributions for the regressors and the errors across segments. This is
important because if an unaccounted variance shift occurs at the same time
inference can be distorted (Pitarakis, 2004). The computation of the robust
version of the Wald test (\ref{robustf}) can be involved. Since the estimate
of $\lambda _{1}$ is $T$-consistent even with correlated errors, an
asymptotically equivalent version is to first take the supremum of the
original Wald test (\ref{ftest}) to obtain the break points, i.e. imposing $%
\Omega =\sigma ^{2}I$. The robust version is obtained by evaluating (\ref%
{robustf}) and (\ref{vhat}) at these estimates, i.e., using $W_{T}^{\ast }(%
\hat{\lambda}_{1};q)$ instead of $\sup_{\lambda _{1}\in \Lambda _{\epsilon
}}W_{T}^{\ast }(\lambda _{1};q)$. An issue of concern for such tests is the
adequacy of the asymptotic distribution as an approximation to the finite
sample distribution. The tests can exhibit size distortions, especially when
the regressors and/or the errors are strongly serially correlated. A
potential solution is to use a bootstrap method (e.g., Prodan, 2008, Chang
and Perron, 2017). Alternatively, given that the estimation of the long-run
variance and, consequently, the choice of the bandwidth, play an essential
role, Cho and Vogelsang (2017) propose a fixed-bandwidth theory. It is shown
to improve upon the standard asymptotic distribution theory, whereby the
bandwidth is negligible relative to $T$. However, while their results are
convincing for a given choice of the bandwidth, when the latter is chosen
endogenously, e.g., using Andrews' (1991)\ method, the improvements are not
as important.

The vast majority of tests considered in the econometrics literature imposes
some trimming that does not consider the possibility of a break occurring
near the beginning or end of the sample. This is not so in the statistics
literature. Such tests lead to a different limiting distribution, mostly
inducing a log-log rate of divergence that needs to be accounted for (e.g.,
Cs\"{o}rg\H{o} and Horv\'{a}th, 1997). These tests usually have poor finite
sample properties. An application for general models in econometrics is
Hidalgo and Seo (2013). Their results are, however, restricted to LM tests
in order to have decent size properties in finite samples.

\noindent \textbf{Optimal tests. }Andrews and Ploberger (1994) consider a
class of tests that maximize a weighted average local asymptotic power
function. They are weighted functions of the standard Wald, LM or LR\
statistics for all permissible break dates. Using either of the three basic
statistics leads to tests that are asymptotically equivalent and we proceed
with the Wald test. Assuming equal weights are given to all break fractions
in some interval $[\epsilon _{1},1-\epsilon _{2}]$, the optimal test for
distant alternatives is the so-called Exp-type test: $Exp$-$W_{T}=\log
(T^{-1}\sum_{T_{1}=[T\epsilon _{1}]+1}^{T-[T\epsilon _{2}]}\exp
((1/2)W_{T}\left( T_{1}/T\right) ))$. For alternatives close to the null
value of no change the optimal test is the $Mean$-$W_{T}$ test: $Mean$-$%
W_{T}=T^{-1}\sum_{T_{1}=[T\epsilon _{1}]+1}^{T-[T\epsilon _{2}]}W_{T}\left(
T_{1}/T\right) $. Kim and Perron (2009) approach the optimality issue from a
different perspective using the approximate Bahadur measure of efficiency.
They show that tests based on the Mean functional are inferior to those
based on the Sup and Exp (which are as efficient) when using the same base
statistic. When considering tests that incorporate a correction for
potential serial correlation in the errors: a) for a given functional, using
the LM statistic leads to tests with zero asymptotic relative efficiency
compared to using the Wald statistic; b) for a given statistic the Mean-type
tests have zero relative efficiency compared to using the Sup and Exp
versions, which are as efficient. Hence, the preferred test should be the
Sup or Exp-Wald tests. Any test based on the LM\ statistic should be
avoided. Such results, and more discussed below, call into question the
usefulness of a local asymptotic criterion to evaluate the properties of
testing procedures; on this issue, see also Deng and Perron (2008).

\noindent \textbf{Non monotonicity in power. }The issue of non-monotonicity
of the power function of structural change tests was first analyzed in
Perron (1991) for changes in a trend function. In more general contexts, the
Sup-Wald and Exp-Wald tests have monotonic power when only one break occurs
under the alternative. As shown in Vogelsang (1999), the Mean-Wald test can
exhibit a non-monotonic power function, though the problem has not been
shown to be severe. All of these, however, suffer from important power
problems when the alternative involves two breaks (Vogelsang, 1997). This
suggests that a test will exhibit a non monotonic power function if the
number of breaks present is greater than the number accounted for. Hence,
though a single break test is consistent against multiple breaks, power
gains can result using tests for multiple structural changes (e.g., the $%
UDmax$ test of BP (1998); see below). Crainiceanu and Vogelsang (2001) also
show how the problem is exacerbated when using estimates of the long-run
variance that allow for correlation. Accordingly, there are problems with
tests based on some local asymptotic arguments (e.g., LM or CUSUM) or that
do no try to model the breaks explicitly. An example is the test of Elliott
and M\"{u}ller (2006) which is deemed optimal for a class of models
involving ``small'' non-constant coefficients. The suggested procedure does
not explicitly model breaks and the test is then akin to a `partial sums
type' test. Perron and Yamamoto (2016) shows that, their test suffers from
severe non-monotonic power, while offering only modest gains for small
breaks. Methods to overcome non-monotonic power problems have been suggested
by Altissimo and Corradi (2003) and Juhl and Xiao (2009). They suggest using
non-parametric methods for the estimation of the mean. The resulting
estimates and tests are, however, very sensitive to the bandwidth used.
There is currently no reliable method to appropriately choose this parameter
in the context of structural changes. Kejriwal (2009) propose to use the
residuals under the alternative to select the bandwidth and those under the
null to compute the long-run variance for the case of a change in mean. Yang
and Vogelsang (2011) show that this can be viewed as an LM test with a
long-run variance constructed with a constrained small bandwidth. They
provide asymptotic critical values based on the fixed-bandwidth asymptotics
of Kiefer and Vogelsang (2005). None of these remedials works when lagged
dependent variables are present.

\noindent \textbf{Tests for multiple structural changes. }A problem with the 
$Mean$-$W_{T}$ and $Exp$-$W_{T}$ tests is that they require computations of
order $O(T^{m})$. Consider instead the Sup-Wald test. With $i.i.d.$ errors,
maximizing the Wald statistic is equivalent to minimizing the SSR, which can
be solved efficiently and the Wald test for $k$ changes is:%
\begin{equation*}
W_{T}(\lambda _{1},...,\lambda _{k};q)=\left( (T-(k+1)q-p)/k\right) \hat{%
\delta}^{\prime }H^{\prime }(H(\bar{Z}^{\prime }M_{X}\bar{Z})^{-1}H^{\prime
})^{-1}H\hat{\delta}/SSR_{k}
\end{equation*}%
with $H$ such that $(H\delta )^{\prime }=(\delta _{1}^{\prime }-\delta
_{2}^{\prime },...,\delta _{k}^{\prime }-\delta _{k+1}^{\prime })$. The
sup-Wald test is $\sup_{(\lambda _{1},...,\lambda _{k})\in \Lambda
_{k,\epsilon }}$ $W_{T}(\lambda _{1},...,\lambda _{k};q)=W_{T}(\hat{\lambda}%
_{1},...,\hat{\lambda}_{k};q)$, where $\Lambda _{\epsilon }=\{(\lambda
_{1},...,\lambda _{k});|\lambda _{i+1}-\lambda _{i}|\geq \epsilon ,\lambda
_{1}\geq \epsilon ,\lambda _{k}\leq 1-\epsilon \}$ and $(\hat{\lambda}%
_{1},...,\hat{\lambda}_{k})=(\hat{T}_{1}/T,...,\hat{T}_{k}/T)$, with $(\hat{T%
}_{1},...,\hat{T}_{k})$ obtained minimizing the SSR over $\Lambda _{\epsilon
}$. With serial correlation and heteroskedasticity in the errors, the test
is 
\begin{equation*}
W_{T}^{\ast }(\lambda _{1},...,\lambda _{k};q)=\left( (T-(k+1)q-p)/k\right) 
\hat{\delta}^{\prime }H^{\prime }(H\hat{V}(\hat{\delta})H^{\prime })^{-1}H%
\hat{\delta},
\end{equation*}%
with $\hat{V}(\hat{\delta})$ as defined by (\ref{vhat}). Again, the
asymptotically equivalent version with the Wald test evaluated at the
estimates $(\hat{\lambda}_{1},...,\hat{\lambda}_{k})$ is used to make the
problem tractable. Critical values are presented in BP (1998) and Bai and
Perron (2003b). The importance of the choice of $\epsilon $ for the size and
power of the test is discussed in BP (2003a) and Bai and Perron (2006).
Often, one may not wish to pre-specify a particular number of breaks. Then a
test of the null hypothesis of no structural break against an unknown number
of breaks given some upper bound $M$ can be used. These are called `double
maximum tests'\textit{. }The first is an equal-weight version defined by $%
UD\max W_{T}(M,q)=\max_{1\leq m\leq M}W_{T}(\hat{\lambda}_{1},...,\hat{%
\lambda}_{m};q)$. The second test applies weights to the individual tests
such that the marginal p-values are equal across values of $m$ denoted $%
WD\max F_{T}(M,q)$ (see BP, 1998). The choice $M=5$ should be sufficient for
most applications. In any event, the critical values vary little as $M$ is
increased beyond $5$. The Double Maximum tests are arguably the most useful
to determine if structural changes are present: 1) there are types of
multiple structural changes that are difficult to detect with a single break
test change (e.g., two breaks with the first and third regimes the same); 2)
the non-monotonic power problem when the number of changes is greater than
specified is alleviated; 3) the power of the double maximum tests is almost
as high as the best power that can be achieved using a test with the correct
number of breaks (BP, 2006).

\noindent \textbf{Sequential tests. }BP (1998)\ also discuss a test of $\ell 
$ versus $\ell +1$ breaks, which can be used to estimate the number of
breaks using a sequential testing procedure. For the model with $\ell $
breaks, the estimated break points denoted by $(\hat{T}_{1},...,\hat{T}%
_{\ell })$ are obtained by a global minimization of the SSR. The strategy
proceeds by testing for the presence of an additional break in each of the $%
(\ell +1)$ segments obtained using the partition $\hat{T}_{1},...,\hat{T}%
_{\ell }$. We reject in favor of a model with $(\ell +1)$ breaks if the
minimal value of the SSR over all segments where an additional break is
included is sufficiently smaller than that from the $\ell $ breaks model.
The break date selected is the one associated with this overall minimum. The
limit distribution of the test is related to that of a test for a single
change. Bai (1999) considers the same problem allowing the breaks to be
global minimizers of the SSR under both the null and alternative hypotheses.
The limit distribution of the test is different. A method to compute the
asymptotic critical values is discussed and the results extended to the case
of trending regressors. These tests can form the basis of a sequential
testing procedure by applying them successively starting from $\ell =0$,
until a non-rejection occurs. The simulation results of BP (2006) show that
such estimate of the number of breaks is better than those obtained using
information criteria as suggested by, e.g., Liu et al. (1997) (see also,
Perron, 1997). But this sequential procedure should not be applied
mechanically. In several cases, it stops too early. The recommendation is to
first use a double maximum test to ascertain if any break is at all present.
The sequential tests can then be used starting at some value greater than $0$%
. Kurozumi and Tuvaandorj (2011) consider useful information criteria
explicitly tailored to structural change problems, which should complement a
sequential testing procedure.

\noindent \textbf{Tests for restricted structural changes. }Consider testing
the null hypothesis of $0$ break versus an alternative with $k$ breaks in a
model which imposes the restrictions $R\delta =r$. In this case, the limit
distribution of the Sup-Wald test depends on the nature of the restrictions
so that it is not possible to tabulate critical values valid in general.
Perron and Qu (2006) discuss a simulation algorithm to compute the relevant
critical values given some restrictions. Imposing valid restrictions results
in tests with much improved power.

\noindent \textbf{Tests for structural changes in multivariate systems. }Bai
et al. (1998) consider a Sup-Wald test for a single common change in a
multivariate system. Qu and Perron (2007) extend the analysis to multiple
structural changes. They consider the case where only a subset of the
coefficients is allowed to change, whether in the parameters of the
conditional mean, the covariance matrix of the errors, or both. The tests
are based on the maximized likelihood ratio over permissible partitions
assuming $i.i.d.$ errors. They can be corrected for serial correlation and
heteroskedasticity when testing for changes in the parameters of the
conditional mean assuming no change in the covariance matrix of the errors.

An advantage of the framework of Qu and Perron (2007) is that it allows
studying changes in the variance of the errors in the presence of
simultaneous changes in the parameters of the conditional mean, thereby
avoiding inference problem when changes in variance are studied in
isolation. Also, it allows for the two types of changes to occur at
different dates, thereby avoiding problems related to tests for changes in
the parameters when a change in variance occurs at some other date. Their
results are, however, only valid in the case of normally distributed errors
when testing for changes in variances (or covariances). This problem was
remedied by Perron and Zhou (2008) who propose tests for changes in the
variances of the errors allowing for changes in the parameters of the
regression in the context of a single equation model. They also consider
various extensions including testing for changes in the parameter allowing
for change in variances and testing for joint changes. These tests are
especially important in light of Hansen's (2000) analysis. First note that
the limit distribution of the tests in a single equation system are valid
under the assumption that the regressors and the variance of the errors have
distributions that are stable across the sample. He shows that when this
condition is not satisfied the limit distribution changes and the test can
be distorted. If the errors are homoskedastic, the size distortions are
quite mild but they can be severe when a change in variance occurs. Both
problems of changes in the distribution of the regressors and the variance
of the errors can be handled using the framework of Qu and Perron (2007) and
Perron and Zhou (2008). If a change in the variance of the residuals is a
concern, one can perform a test for no change in some parameters of the
conditional model allowing for a change in variance. If changes in the
marginal distribution of some regressors is a concern, one can use a
multi-equations system with equations for these regressors.

\noindent \textbf{Tests valid with }$\mathbf{I(1)}$\textbf{\ regressors. }%
With $I(1)$ regressors, a case of interest is a system of cointegrated
variables. For testing, Hansen (1992) considers the null hypothesis of no
change in both coefficients and proposed Sup and Mean-LM tests for a one
time change. He also considers a version of the LM test directed against the
alternative that the coefficients are random walk processes. Kejriwal and
Perron (2010a) provide a comprehensive treatment of issues related to
testing for multiple structural changes at unknown dates in cointegrated
regression models using the Sup-Wald test. They allow both $I(0)$ and $I(1)$
variables and derive the limiting distribution of the Sup-Wald test for a
given number of cointegrating regimes. They also consider the double maximum
tests and provide critical values for a wide variety of models that are
expected to be relevant in practice. The asymptotic results have important
implications for inference. It is shown that in models involving both $I(1)$
and $I(0)$ variables, inference is possible as long as the intercept is
allowed to change across regimes. Otherwise, the limiting distributions of
the tests depend on nuisance parameters. They propose a modified Sup-Wald
test that has good size and power properties. Note, however, that the Sup
and Mean-Wald test will also reject when no structural change is present and
the system is not cointegrated. Hence, the application of such tests should
be interpreted with caution. No test is available for the null hypothesis of
no change in the coefficients allowing the errors to be $I(0)$ or $I(1)$.
This is because when the errors are $I(1)$, we have a spurious regression
and the parameters are not identified. To be able to properly interpret the
tests, they should be used in conjunction with tests for the presence or
absence of cointegration allowing shifts in the coefficients (see, Perron,
2006).

\noindent \textbf{Tests valid whether the errors are }$\mathbf{I(1)}$\textbf{%
\ or }$\mathbf{I(0)}$\textbf{. }The issue of testing for structural changes
in a linear model with errors that are either $I(0)$ or $I(1)$ is of
interest when the regression is a polynomial time trend (e.g., testing for a
change in the slope of a linear trend). The problem is to devise a procedure
that has the same limit distribution in both the $I(0)$ and $I(1)$ cases.
The first to provide such a solution is Vogelsang (2001). He also accounts \
for correlation with an autoregressive approximation so that the Wald test
has a non-degenerate limit distribution in both the $I(0)$ and $I(1)$ cases.
The novelty is that he weights the statistic by a unit root test scaled by
some parameter. For any given significance level, a value of this scaling
parameter can be chosen so that the asymptotic critical values will be the
same. His simulations show, however, the test to have little power in the $%
I(1)$ case so that he resorts to advocating the joint use of that test and a
normalized Wald test that has good properties in the $I(1)$ case but has
otherwise very little power in the $I(0)$ case. Perron and Yabu (2009b)\ and
Harvey et al. (2009) have independently proposed procedures that achieve the
same goal and that were shown to have better size and power than that of
Vogelsang (2001). The approach of Harvey et al. (2009) builds on the work of
Harvey et al. (2007). It is based on a weighted average of the regression
t-statistic for a change in the slope of the trend appropriate for the case
of $I(0)$ and $I(1)$ errors. In the former case a regression in levels is
used while in the latter a regression in first-differences is used. With an
unknown break date, the supremum over a range of possible break dates is
taken. As in Vogelsang (2001), a correction is required to ensure that, for
a given significance level, the weighted test has the same asymptotic
critical value in both the $I(0)$ and $I(1)$ cases.

Perron and Yabu (2009b) builds on Perron and Yabu (2009a) who analyzed
hypothesis testing on the slope coefficient of a linear trend model. The
method is based on a Feasible Quasi Generalized Least Squares approach that
uses a superefficient estimate of the sum of the autoregressive parameters $%
\alpha $ when $\alpha =1$. The estimate of $\alpha $ is the OLS estimate
from an autoregression applied to detrended data and is truncated to take a
value $1$ whenever it is in a $T^{-\delta }$ neighborhood of $1$. This makes
the estimate ``super-efficient''\ when $\alpha =1$. Theoretical arguments
and simulation evidence show that $\delta =1/2$ is the appropriate choice.
Perron and Yabu (2009b) analyze the case of testing for changes in level or
slope of the trend function of a univariate time series. When the break
dates are unknown, the limit distribution is nearly the same in the $I(0)$
and $I(1)$ cases using the Exp-Wlad test. Hence, it is possible to have
tests with nearly the same size in both cases. To improve the finite sample
properties, use is made of a bias-corrected version of the OLS estimate. The
Perron-Yabu test has greater power overall; see Chun and Perron (2013).
Kejriwal and Perron (2010b) extend the results to show that the test of
Perron and Yabu (2009b) can be applied in a sequential manner using the same
critical values. An alternative perspective was provided by Sayginsoy and
Vogelsang (2011) who use a fixed-bandwith asymptotic theory. Extensions that
allow the errors to be fractionally integrated have been considered by
Iacone et al. (2013a,b).

\noindent \textbf{Testing for common breaks}. Oka and Perron (2017) consider
testing for common breaks across or within equations in a multivariate
system. The framework is very general and allows stationary or integrated
regressors and trends. The null hypothesis is that breaks in different
parameters occur at common locations or are separated by some positive
fraction of the sample size. Under the alternative hypothesis, the break
dates are not the same and need not be separated by a positive fraction of
the sample size across parameters. A quasi-likelihood ratio test assuming
normal errors is used. The quantiles of the limit distribution need to be
simulated and an efficient algorithm is provided. Kim, Oka, Estrada and
Perron (2017) extend this work to cover \ the case of testing for common
breaks in a system of equations involving joint-segmented trends. The
motivation was spurred by a need to test whether the breaks in the slope of
the trend functions of temperatures and radiative forcing, occurring in
1960, are common (see, Estrada, Perron, Martinez-Lopez, 2013).

\noindent \textbf{Band spectral regressions and low frequency changes}.
Perron and Yamamoto (2013) consider the issue of testing for structural
change using a band-spectral analysis. They allow changes over time within
some frequency bands, permitting the coefficients to be different across
frequency bands. Using standard assumptions, the limit distributions
obtained are similar to those in the time domain counterpart. They show that
when the coefficients change only within some frequency band (e.g., the
business cycle) we can have increased efficiency of the estimates of the
break dates and increased power for the tests provided, of course, that the
user chosen band contains the band at which the changes occur. They also
discuss a very useful application in which the data is contaminated by some
low frequency process and that the researcher is interested in whether the
original non-contaminated model is stable. For example, the dependent
variable may be affected by some random level shift process (a low frequency
contamination) but at the business cycle frequency the model of interest is
otherwise stable. They show that all that is needed to obtain estimates of
the break dates and tests for structural changes that are not affected by
such low frequency contaminations is to truncate a low frequency band that
shrinks to zero at rate $\log (T)/T$. Simulations show that the tests have
good sizes for a wide range of truncations. The exact truncation does not
really matter, as long as some of the very low frequencies are excluded.
Hence, the method is quite robust. They also show that the method delivers
more precise estimates of the break dates and tests with better power
compared to using filtered series via a band-pass filter or with a
Hodrick-Prescott (1997) filter. This work is related to a recent strand in
the literature that attempts to deliver methods robust to low frequency
contaminations. One example pertains to estimation of the long-memory
parameter. It is by now well known that spurious long-memory can be induced
by level shifts or various kinds of low frequency contaminations. Perron and
Qu (2007, 2010), Iacone (2010), McCloskey and Perron (2013) and Hou and
Perron (2015) exploit the fact low frequency contaminations will produce
peaks in the periodograms at a very few low frequencies, and suggest robust
procedures eliminating such low frequencies. Tests for spurious versus
genuine long-memory have been proposed by Qu (2011). McCloskey and Hill
(2017) provide a method applicable to various time series models, such as
ARMA, GARCH and stochastic volatility models.

\noindent \textbf{Endogenous regressors}. Consider a model with errors
correlated with the regressors: 
\begin{equation}
y=\bar{X}\delta +u,  \label{dgp}
\end{equation}%
where $\bar{X}=diag(X_{1},...,X_{m+1})$, a $T$ by $(m+1)p$ matrix with $%
X_{i}=(x_{T_{i-1}+1},...,x_{T_{i}})^{\prime }$ ($i=1,...,m+1$). This
includes partial structural change models imposing $R\delta =r$ with $R$ a $%
k $ by $(m+1)p$ matrix. When the regressors are correlated with the errors,
we assume a set of $q$ variables $z_{t}$ that can serve as instruments, and
define the $T$ by $q$ matrix $Z=(z_{1},...,z_{T})^{\prime }$. We consider a
reduced form linking $Z$ and $X$ that itself exhibits $m_{z}$ changes, so
that $X=\bar{W}^{0}\theta ^{0}+v$, with $\bar{W}%
^{0}=diag(W_{1}^{0},...,W_{m_{z}+1}^{0})$, the diagonal partition of $W$ at
the break dates $(T_{1}^{z0},...,T_{m_{z}}^{z0})$ and $\theta ^{0}=(\theta
_{1}^{0},...,\theta _{m_{z}+1}^{0})$. Also, $v=(v_{1},...,v_{T})^{\prime }$
is a $T$ by $q$ matrix, which can be correlated with $u_{t}$ but not with $%
z_{t}$. Given estimates $(\hat{T}_{1}^{z},...,\hat{T}_{m_{z}}^{z})$ obtained
using the method of BP (2003a), one constructs $\hat{W}$ $=diag(\hat{W}%
_{1},...,\hat{W}_{m_{z+1}})$, a $T$ by $(m_{z}+1)q$ matrix with $\hat{W}%
_{l}=(w_{\hat{T}_{l-1}^{z}+1},...,w_{\hat{T}_{l}^{z}})^{\prime }$ \ for $%
l=1,...,m_{z}+1$. Let $\hat{\theta}$ be the OLS estimate in a regression of $%
X$ on $\hat{W}$. The instruments are $\hat{X}=\hat{W}\hat{\theta}=diag(\hat{X%
}_{1}^{\prime },...,\hat{X}_{m_{z}+1}^{\prime })^{\prime }$ where $\hat{X}%
_{l}=\hat{W}_{l}(\hat{W}_{l}^{\prime }\hat{W}_{l})^{-1}\hat{W}_{l}^{\prime }%
\tilde{X}_{l}$ with $\tilde{X}_{l}=(x_{\hat{T}_{l-1}^{z}+1},...,x_{\hat{T}%
_{l}^{z}})^{\prime }$. The instrumental variable (IV) regression is%
\begin{equation}
y=\bar{X}^{\ast }\delta +\tilde{u},  \label{2sls}
\end{equation}%
subject to the restrictions $R\delta =r$, where $\bar{X}^{\ast }=diag(\hat{X}%
_{1},...,\hat{X}_{m+1})$, a $T$ by $(m+1)p$ matrix with $\hat{X}_{j}=(\hat{x}%
_{T_{j-1}+1},...,\hat{x}_{T_{j}})^{\prime }$ ($j=1,...,m+1$). Also, $%
\tilde{u}=(\tilde{u}_{1},...,\tilde{u}_{T})^{\prime }$ with $\tilde{u}%
_{t}=u_{t}+\eta _{t}$ where $\eta _{t}=(x_{t}^{\prime }-\hat{x}_{t}^{\prime
})\delta _{j}$ for $T_{j-1}^{0}+1\leq t\leq T_{j}^{0}$. The estimates of the
break dates are $(\hat{T}_{1},...,\hat{T}_{m_{x}})=\arg
\min_{T_{1},...,T_{m_{x}}}SSR_{T}^{R}(T_{1},...,T_{m})$, where $%
SSR_{T}^{R}(T_{1},...,T_{m})$ is the SSR from (\ref{2sls}) evaluated at $%
\{T_{1},...,T_{m}\}$. Perron and Yamamoto (2014) provide a simple proof of
the consistency and limit distribution of the estimates of the break dates
showing that using generated (or second stage) regressors implies that the
assumptions of Perron and Qu (2006) are satisfied. For an earlier, more
elaborate though less comprehensive treatment, see Hall et al. (2012) and
Boldea et al. (2012). Hence, all results of BP (1998) carry through, but
care must be applied when the structural and reduced forms contain
non-common break.

Of substantive interest is that the IV approach is not necessary as
discussed in Perron and Yamamoto (2015); one can simply still use OLS
applied to (\ref{dgp}). First, except for a knife-edge case, changes in the
true parameters imply a change in the probability limits of the OLS
estimates, which is equivalent in the leading case of regressors and errors
having a homogenous distribution across segments. Second, one can
reformulate the model with those limits as the basic parameters so that the
regressors and errors are uncorrelated. We are then back to the standard
framework. Importantly, using OLS involves the original regressors while IV
the second stage regressors, which have less quadratic variation since $%
||P_{Z}X||\leq ||X||$. Hence, in most cases, a given change in the
parameters will cause a larger change in the conditional mean of the
dependent variable using OLS compared with IV. It follows that using OLS
delivers consistent estimates of the break fractions and tests with the
usual limit distributions and also improves on the efficiency of the
estimates and the power of the tests in most cases. Also, OLS avoids weak
identification problems inherent when using IV\ methods. Some care must,
however, be exercised. Upon a rejection, one should verify that the change
in the probability limit of the OLS parameter estimates is not due to a
change in the bias terms. In most applications, there will be no change in
the bias but still one should be careful to assess the source of the
rejection. This is easily done since after obtaining the OLS-based estimates
of the break dates one would estimate the structural model based on such
estimates. The relevant quantities needed to compute the change in bias
across segments are then directly available. To elaborate, assume known
break dates and let $p\lim_{T\rightarrow \infty }(\Delta
T_{i}^{0})^{-1}\tsum\nolimits_{t=T_{i-1}^{0}+1}^{T_{i-1}^{0}+[s\Delta
T_{i}^{0}]}x_{t}x_{t}^{\prime }=Q_{XX}^{i}$ and $p\lim_{T\rightarrow \infty
}E(X_{i}u)=\phi _{i}$ ($i=1,...,m+1$), the probability limit of the OLS
estimate is $\delta ^{\ast }=\delta ^{0}+[(Q_{XX}^{1})^{-1}\phi
_{1},...,(Q_{XX}^{m+1})^{-1}\phi _{m+1})]^{\prime }$. Any change in $\delta
^{0}$ imply a change in $\delta ^{\ast }$, except for a knife-edge case when
the change in the bias exactly offsets the change in $\delta ^{0}$. Hence,
one can still identify parameter changes using OLS and a change in $\delta
^{0}$ will, in general, cause a larger change in the conditional mean of $%
y_{t}$. Consider writing (\ref{dgp}) as%
\begin{equation*}
y=\bar{X}_{0}\delta ^{0}+P_{\bar{X}_{0}}u+(I-P_{\bar{X}_{0}})u=\bar{X}%
_{0}[\delta ^{0}+(\bar{X}_{0}^{\prime }\bar{X}_{0})^{-1}\bar{X}_{0}^{\prime
}u]+(I-P_{\bar{X}_{0}})u=\bar{X}_{0}\delta _{T}^{\ast }+u^{\ast },
\end{equation*}%
where $u^{\ast }=(I-P_{\bar{X}_{0}})u$ and $\delta _{T}^{\ast }=[\delta
^{0}+(\bar{X}_{0}^{\prime }\bar{X}_{0})^{-1}\bar{X}_{0}^{\prime }u]$ for
which $\delta _{T}^{\ast }\rightarrow _{p}\delta ^{\ast }$. So we can
consider a regression in terms of the population value of the parameters.
Now, $\bar{X}_{0}$ is uncorrelated with $u^{\ast }$ so that the OLS
estimate, say $\hat{\delta}^{\ast }$, is consistent for $\delta ^{\ast }$.
This suggests estimating the break dates by minimizing the SSR from the
regression $y=\bar{X}\delta ^{\ast }+u^{\ast }$. OLS dominates IV except for
a narrow case, unlikely in practice. The loss in efficiency when using 2SLS
can be especially pronounced when the instruments are weak as is often the
case. Of course, the ultimate goal is not to get estimates of the break
dates per se but of the parameters within each regime, one should then use
an IV regression but conditioning on the estimates of the break dates
obtained using the OLS-based procedure. Their limit distributions will, as
usual, be the same as if the break dates were known. Using the same logic,
tests for structural change are more powerful when based on OLS rather than
IV. This idea was used by Kurozumi (2017) to show that, with endogenous
regressors, using OLS is better when monitoring online for a structural
change using a CUSUM-type method.

\noindent \textbf{Quantile Regressions}. Following Oka and Qu (2011), assume
that the $\tau $th conditional quantile function of $y_{t}$ given $z_{t}$ is
linear in the parameters and given by $Q_{y_{t}}(\tau
|z_{t})=F_{y_{t}|z_{t}}^{-1}(\tau |z_{t})=z_{t}\beta (\tau )$. The
population coefficient $\beta (\tau )$ is known to be the minimizer of the
criterion function $Q_{\tau }(\beta )=E[\rho _{\tau }(y-z^{\prime }\beta )]$
where $\rho _{\tau }(a)$ is the check function given by $\rho _{\tau
}(a)=a(\tau -1(a<0))$; see Koenker (2005). Without structural changes, the
quantile regression estimator of $\beta (\tau )$ is the minimizer of the
empirical analog $\hat{Q}_{\tau }(\beta )=\sum_{t=1}^{T}\rho _{\tau
}(y_{t}-z_{t}^{\prime }\beta )$. Suppose that the $\tau $th quantile has $m$
structural changes, occurring at unknown dates $(T_{1},...,T_{m})$, such that%
\begin{equation}
Q_{y_{t}}(\tau |z_{t})=z_{t}^{\prime }\theta _{j}(\tau ),\quad
t=T_{j-1}+1,...,T_{j}  \label{eq:1}
\end{equation}%
for $j=1,...,m+1$. The vector $\theta _{j}(\tau )$ are the quantile
dependent unknown parameters, with possible restrictions to allow for
partial structural changes. Qu (2008) proposes a fluctuation type statistic
based on the subgradient and a Wald type statistic based on comparing
parameter estimates from different subsamples. They can be used to test for
changes occurring in a prespecified quantile, or across quantiles. The
limiting distributions under the null is nuisance parameter free and can be
simulated. Oka and Qu (2011) consider the estimation of multiple structural
changes at unknown dates in one or multiple conditional quantile functions.
A procedure to determine the number of breaks is also discussed. The method
can deliver more informative results than the analysis of the conditional
mean function alone.

\noindent \textbf{Lasso}. A growing literature uses Lasso-type methods to
address structural change problems, which can estimate the location and
number of breaks simultaneously. Estimating structural changes can be viewed
as a variable selection problem and Lasso estimates the regression
coefficients by minimizing the usual SSR with a penalty for model complexity
through the sum of the absolute values of the coefficients. Assume, for
simplicity, $p=0$ and $q=1$, i.e., a pure structural change model with a
single regressor. When $m=1$, the model is:%
\begin{equation*}
y_{t}=z_{t}\delta _{1}+(\delta _{2}-\delta _{1})z_{t}\mathbf{1}(t\geq
T_{1}+1)+u_{t}.
\end{equation*}%
When there is a break at $t=T_{1}$, then $\delta _{2}\neq \delta _{1}\neq 0$%
, otherwise $\delta _{2}=\delta _{1}$. Using a set of regressors $%
g(z_{t})=\{z_{t}\mathbf{1}(t\geq \tilde{t}),\forall \tilde{t}\in \lbrack 
\underline{t},\bar{t}],1<\underline{t}<\bar{t}<T\}$, we can express the
model as:%
\begin{eqnarray*}
y_{t} &=&z_{t}\delta _{1}+b_{1}z_{t}\mathbf{1}(t\geq \underline{t})+...+b_{%
\bar{t}-\underline{t}}z\mathbf{1}(t\geq \bar{t})+u_{t} \\
&=&\underset{w_{0,t}}{\underbrace{z_{t}}}\delta _{1}+\underset{w_{1,t}}{(%
\underbrace{z_{t}\mathbf{1}(t\geq \underline{t})}})b_{1}+...+\underset{%
w_{(T_{1}-\underline{t}+2),t}}{(\underbrace{z_{t}\mathbf{1}(t\geq T_{1}+1)})}%
\underset{\delta _{2}-\delta _{1}}{\underbrace{b_{T_{1}-\underline{t}+2}}}%
+...+(\underset{z_{(\bar{t}-\underline{t}+1),t}}{\underbrace{z_{t}\mathbf{1}%
(t\geq \bar{t})}})b_{\bar{t}-\underline{t}+1}+u_{t} \\
&=&w_{0,t}\theta _{0}+g(w_{0,t})\theta _{1}+u_{t}=w_{t}^{\prime }\theta
+u_{t}
\end{eqnarray*}%
with $b_{i}=0$, $i=1,...,T_{1}-\underline{t}+1,T_{1}-\underline{t}+3,...,%
\bar{t}-\underline{t}+1$ and $b_{T_{1}-\underline{t}+2}=|\delta _{2}-\delta
_{1}|\neq 0$. Denote by $g(x_{0,t})=\{w_{1,t},...,w_{(\bar{t}-\underline{t}%
+1),t}\}$ the transformed regressors generated from $w_{0,t}=z_{t}$ whose
coefficient is subject to change. Then $w_{t}=\{w_{0,t},g(w_{0,t})\}$ is the
complete set of regressors. If we can consistently estimate the coefficients
associated with $g(x_{0,t})$ that are greater than zero, we can date the
break point. OLS would not provide consistent estimates because the number
of regressors is too large. The method is flexible; e.g., if one has some
prior knowledge that a change has occurred at some date, dummy variable can
be added without the associated generated regressors. A model with multiple
structural changes in many regressors can be obtained as an extension. One
simply let $w_{0,t}$ be the vector of regressors whose coefficients are
subject to change and $g(w_{0,t})$ be the vector of artificially constructed
regressors obtained from the original ones. In general, the number of
regressors is $n=p+q(\ddot{r}+1)$, where $\ddot{r}=\bar{t}-\underline{t}+1$
is the number of transformed regressors associated with the original
regressors whose coefficients are allowed to change (fewer are possible if
there is prior information about where the breaks cannot occur). Hence, $n$
can be very large, much larger than $T$. The structural break model has a
sparse pattern since few coefficients are non-zero, namely $s=p+q(m+1)$. The
Lasso estimator for sparse models is $\hat{\theta}=\arg \min_{\theta \in 
\mathbb{R}^{q}}\hat{Q}(\theta )+(\lambda /T)\Vert \hat{\Upsilon}\theta \Vert
_{1}$, where $\lambda $ is the penalty level, $\hat{Q}(\theta
)=\sum_{t=1}^{T}(y_{t}-w_{t}^{\prime }\theta )^{2}$, $\hat{\Upsilon}=diag(%
\hat{\gamma}_{1},...,\hat{\gamma}_{p})$ is a diagonal matrix with the
penalty loadings and $\Vert \hat{\Upsilon}\theta \Vert _{1}=\sum_{j=1}^{p}|%
\hat{\gamma}_{j}\theta _{j}|$ is the $\ell _{1}$-norm. Ideally, these are
adapted to information about the error term, which is not feasible since $%
u_{t}$ is not observed. In practice, we can use the estimated residuals and
proceed via iterations or simply assume homoskedastic Gaussian errors, in
which case $\hat{\Upsilon}$ is the identity matrix. Often, additional
thresholding is applied to remove regressors with small estimated
coefficients which may have been included due to estimation error. Then, the
thresholded Lasso estimator is $\hat{\theta}(t_{L})=(\hat{\theta}_{j}1\{|%
\hat{\theta}_{j}|>t_{L}\},j=1,...,q)$ where $t_{L}\geq 0$ is the threshold
level. The problem is in the choice of $\lambda $ and $t_{L}$. Results are
well established for a random sample of data. With serially correleated
series, things are more complex.

The following is a partial list of some relevant papers using Lasso for
models with structural changes. Harchaoui and L\'{e}vy-Leduc (2010) propose
a total variation penalty to estimate changes in the mean of a sequence of
independent and identically distributed (\textit{i.i.d.}) Gaussian random
variables. Bleakley and Vert (2011) propose a group fused Lasso method for
changes in the mean of a vector of \textit{i.i.d.} Gaussian random variables
assumed to share common break points. Chan, Yau, and Zhang (2014) consider
using a group Lasso method for changes in an autoregressive model with
heteroskedastic errors. They suggest a two steps method involving the use of
an information criterion to select the number of break points. Ciuperca
(2014) considers multiple changes in a linear regression model with i.i.d.
errors using Lasso with an information criterion or adaptive Lasso. Rojas
and Wahlberg (2014) use penalized maximum likelihood estimator for changes
in the mean of a sequence of \textit{i.i.d.} Gaussian random variables. Aue,
Cheung, Lee and Zhong (2014) consider structural breaks in conditional
quantiles using the minimum description length principle. While the
framework is quite general, the method cannot consistently estimate the
number of breaks jointly with their location. Qian and Su (2016) consider
estimation and inference of common breaks in panel data models with
endogenous regressors. The regressors and errors are, however, restricted to
be \textit{i.i.d.} processes, a common feature in this literature up to now.
Allowing for general mixing regressors and errors has not, to our knowledge,
been achieved. Work is needed to achieve the level of generality available
using standard procedures. Nevertheless, it remains a promising approach,
especially in the context of very large datasets.

\noindent \textbf{Factors.} The issue of structural breaks in factor models
has recently received considerable attention; for a more detailed survey see
Bai and Han (2016). We first discuss in some details the methods of Baltagi,
Kao and Wang (2017) and Bai, Han and Shi (2017) and then briefly mention
other works. The high dimensional factor model with $m$ changes in the
factor loadings considered by Baltagi, Kao and Wang (2017) is given by 
\begin{equation*}
y_{it}=f_{t}^{\prime }\phi _{i}+f_{b,t}^{\prime }\phi _{i,j}+u_{it},\qquad
\qquad (t=T_{j-1}^{0}+1,\ldots ,\,T_{j}^{0})
\end{equation*}%
for $j=1,\ldots ,\,m+1,\,i=1,\ldots ,\,N$ and $t=1,\ldots ,\,T$ where $f_{t}$
and $f_{b,t}$ are vectors of factors without and with changes in the
loadings, respectively, $\phi _{i}$ and $\phi _{i,j}$ are the factor
loadings of unit $i$ corresponding to $f_{t}$ and $f_{b,t}$ in the $j$-th
regime, respectively, and $u_{it}$ is a disturbance term which can have
serial and cross-sectional dependence as well as heteroskedasticity. The
problem is to estimate the break points, determine the number of factors,
and estimate the factors and loadings in each regime. We first discuss a
procedure when the number of breaks is known. It first estimates the break
points using a simultaneous or sequential method, which leads to consistent
estimates for $\lambda _{b}^{0}=T_{b}^{0}/T$, though not for $T_{b}^{0}$.
Secondly, it involves plugging-in the break points estimates and estimating
the number of factors and the factor space in each regime. Since the factors
are latent, one has to determine the number of pseudo factors which is akin
to selecting moment conditions whereas in BP (1998) the model is parametric
and the moment conditions are known a priori. Baltagi, Kao and Wang (2017)
propose to convert the statistical problem from estimating multiple changes
in the loadings to estimating changes in the pseudo factors. This relies on
the fact that the mean of the second moment matrix of the pseudo factors
have changes at the same dates as the loadings. After this conversion, the
data become fixed dimensional with observable regressors and conceptually
the problem becomes similar to that of Qu and Perron (2007).

Estimation of the break points involve three steps: (1) ignoring breaks, use
any consistent estimator $\tilde{r}$ to estimate the number of factors; (2)
estimate the first $\tilde{r}$ factors $\tilde{g}_{t}$ using the Principal
Component (PC) method; 3) for any partition $\left( T_{1},\ldots
,\,T_{m}\right) $, split the sample into $m+1$ subsamples, estimate $\tilde{%
\Sigma}_{j}=\left( T_{j}-T_{j-1}\right) ^{-1}\sum_{t=T_{j-1}+1}^{T_{j}}%
\tilde{g}_{t}\tilde{g}_{t}^{\prime }$ and calculate the SSR, 
\begin{equation*}
\tilde{S}\left( T_{1},\ldots ,\,T_{m}\right)
=\tsum\nolimits_{j=1}^{m+1}\tsum\nolimits_{t=T_{j-1}+1}^{T_{j}}[\mathrm{vech}%
(\tilde{g}_{t}\tilde{g}_{t}^{\prime }-\tilde{\Sigma}_{j})]^{\prime }[\mathrm{%
vech}(\tilde{g}_{t}\tilde{g}_{t}^{\prime }-\tilde{\Sigma}_{j})].
\end{equation*}%
The estimates of the break points are the minimizer of $\tilde{S}\left(
T_{1},\ldots ,\,T_{m}\right) $. The motivation is that the second moment
matrix of $g_{t}$ has breaks at the same dates as the factor loadings. The
results are valid under general high-level conditions on the errors and the
factors are allowed to be dynamic and to include lags. To estimate the
number of factors, these can be correlated with the errors. The limiting
distributions of $\hat{T}_{b}-T_{b}^{0}$ $\left( b=1,\ldots ,\,m\right) $
have the same form as the one for the single break case derived in Baltagi,
Kao and Wang (2016).

Consider now testing for multiple changes in the factor loadings. Following
BP (1998), Baltagi, Kao and Wang (2017) propose two different tests: no
change versus a fixed number of changes; $l$ versus $l+1$ changes. The limit
distributions follow those in BP (1998). The first test loses power when the
number of changes is mispecified. They propose adapting the UDmax and WDmax
tests of BP (1998) allowing an unknown number of changes (up to some upper
bound). For to the test of $m=l$ versus $m=l+1,$ one first estimates the
break points and once they are plugged in, testing for $m=l$ versus $m=l+1$
changes becomes equivalent to testing no change versus a single change in
each regime jointly. The null limiting distribution is obtained by
simulations and depends on the number of factors in each regime. When this
number is stable, it is similar to that in BP (1998).

Bai, Han and Shi (2017) study the properties of the least-squares estimator
of the single break point in a high dimensional factor model. The model is
given by%
\begin{equation*}
y_{it}=\phi _{ij}^{\prime }f_{t}+u_{it},\text{for\textrm{\ }}\mathrm{\,}%
t=1,\ldots ,\,T_{b}^{0}\text{, if }j=1\text{ and }t=T_{b}^{0}+1,\ldots ,\,T%
\text{, if }j=2\text{,}
\end{equation*}%
for $i=1,\ldots ,\,N$ where $T_{b}^{0}=T\lambda _{b}^{0}$ is the unknown
common break point. The estimation of $T_{b}^{0}$ involves the estimated
latent factors. The model in matrix format is: 
\begin{equation}
\begin{bmatrix}
Y_{T_{b}^{0}}^{\left( 1\right) } \\ 
Y_{T_{b}^{0}}^{\left( 2\right) }%
\end{bmatrix}%
=%
\begin{bmatrix}
F^{\left( 1\right) } & 0_{T_{b}^{0}\times r} \\ 
0_{\left( T-T_{b}^{0}\right) \times r} & F^{\left( 2\right) }%
\end{bmatrix}%
\begin{bmatrix}
\Lambda _{1}^{\prime } \\ 
\Lambda _{2}^{\prime }%
\end{bmatrix}%
+%
\begin{bmatrix}
u_{T_{b}^{0}}^{\left( 1\right) } \\ 
u_{T_{b}^{0}}^{\left( 2\right) }%
\end{bmatrix}%
\equiv G\Theta ^{\prime }+u  \label{Model Factor Bai 2}
\end{equation}%
Model \eqref{Model Factor Bai 2} is an observationally equivalent factor
model, where the number of factors is doubled and the factor loadings are
time invariant. Under \eqref{Model Factor Bai 2}, the factor process has
structural breaks, which was the basis for the framework of Baltagi, Kao and
Wang (2017). The estimates of the break points are the minimizer of 
\begin{equation*}
SSR(T_{b},\,\tilde{F})=\tsum\nolimits_{i=1}^{N}\tsum%
\nolimits_{t=1}^{T_{b}}(y_{it}-\tilde{f}_{t}^{\prime }\tilde{\phi}%
_{i1})^{2}+\tsum\nolimits_{i=1}^{N}\tsum\nolimits_{t=T_{b}+1}^{T}(y_{it}-%
\tilde{f}_{t}^{\prime }\tilde{\phi}_{i2})^{2}.
\end{equation*}%
over $T\lambda _{1}\leq T_{b}\leq T\lambda _{2}$, where $\tilde{f}$ are
estimates of $f_{t}$ using a principal component method. Since the factors
are estimated by a PC method, they are not efficient. However, the
efficiency loss relative to the maximum likelihood estimator, vanishes as $%
N,\,T\rightarrow \infty $. The errors are assumed to be independent from the
factors and loadings, however, $u_{it}$ can be weakly correlated in both
cross-sectional and time dimensions. There can be dependence between $%
\left\{ f_{t}\right\} $ and $\left\{ \phi _{i1},\,\phi _{i2}\right\} $ and
the break magnitudes $\phi _{i2}-\phi _{i1}$ can be dependent on the
factors. Theoretical results are provided under both large and small breaks
where the latter are modeled in two ways: 1) the magnitude of the change in
each factor loading is of order $N^{\frac{\nu -1}{2}}$ for some $0<\nu \leq
1 $; 2) the magnitude of the change is fixed but only $O\left( N^{\nu
}\right) $ units have a break for some $0<\nu \leq 1$. The case $\nu =1$
corresponds to large breaks studied by Chen, Dolado and Gonzalo (2014), Han
and Inoue (2015), Cheng, Liao and Schrfheide (2015), and Baltagi, Kao and
Wang (2015, 2017). Discussing separately large and small breaks is useful
given the asymptotic properties of the PC estimator of the factors. For
small breaks, $\tilde{T}_{b}$ is consistent for $T_{b}^{0}$ as $%
N,\,T\rightarrow \infty $ provided conditions on the ratio $N/T$ are
satisfied. The consistency result for $\tilde{T}_{b}$ is strong and
different from the univariate case for which only the break fraction is
consistently estimated. For large breaks, $T_{b}^{0}$ is not consistently
estimable. The framework under small breaks allows a non-degenerate
asymptotic distribution similar to that of the OLS break point estimator in
panel models (cf. Bai, 2010, and the section on panels below). It does not
depend on the exact distribution of the errors but depends on the
distribution of the factors $f_{t}$, being the same whether $f_{t}$ is
observable or not. Hence, evaluating the limit distribution using the
plug-in approach, replacing population quantities with consistent estimates,
is not applicable due to rotations. Bai, Han and Shi (2017) propose a
bootstrap method, which, however, lacks robustness to cross-sectional
correlation.

Additional remarks on factor models follow. First, Bai, Han and Shi's (2017)
analysis assumes a known number of factors. This is a strong restriction
which future research should relax. Su and Wang (2017) developed an
innovative adaptive group Lasso estimator that can determine the number of
factors and the break fraction simultaneously but it is valid only under
large breaks in the loading matrix. Therefore, joint estimation of the break
points and the number of factors remains an open issue even from a
computational perspective. It is also necessary to consider alternative
inference methods for the break points because the bootstrap procedure does
not work very well for small breaks. Other authors have proposed alternative
methods for estimating factor models with breaks. Cheng, Liao and
Schorfheide (2016) developed a shrinkage method that can consistently
estimate the break fraction. Chen (2015) considers a least-squares estimator
of the break points and proves the consistency of estimated break fractions
while Massacci (2015) studies the least-squares estimation of structural
changes in factor loadings under a threshold model setup. Additional tests
for structural changes in factor models have been proposed; see Chen, Dolado
and Gonzalo (2014), Corradi and Swanson (2014), Han and Inoue (2015), Su and
Wang (2015) and Cheng, Liao and Schorfheide (2016). Breitung and Eickmeier
(2011) propose a test for dynamic factor models. Yamamoto and Tanaka (2015)
show that their test has nonmonotonic power and propose a modified version
that solves the problem. The major restriction for most studies is their
focus on testing for a common break date in the factor loadings. While a
common break date is sometimes relevant, one cannot exclude the possibility
that some of the loadings have breaks at different dates. Additional work is
needed in that direction.

\noindent \textbf{Panels.} Panel data studies have become increasingly
popular including inference about breaks. The literature on estimating panel
structural breaks can be categorized as assuming whether the parameters of
interest are allowed to be heterogenous across units or not. We focus on
heterogeneous panels since they are more relevant in practice and refer the
reader to De Watcher and Tzavalis (2012) and Qian and Su (2014) for
corresponding methods for homogeneous panels. Bai (2010) considers the
problem of estimating a common break point in a panel with $N$ units and $T$
observations for each unit. The model takes the form: $y_{it}=\mu
_{ij}+u_{it}$, for\textrm{\ }$\mathrm{\,}t=1,\,\ldots ,\,T_{b}^{0}$, if $j=1$
and $t=T_{b}^{0}+1,\ldots ,\,T$, if $j=2$, where $i=1,\,\ldots ,\,N$, and $%
u_{it}$ is a disturbance term. The common break specification means that
each unit has a break point at $T_{b}^{0}$. The model allows for
heterogeneous means and break magnitudes $\mu _{i2}-\mu _{i1}$. Bai (2010)
provides results for both fixed $T$ and $T$ going to infinity. Unlike in the
univariate model where the break point is assumed to correspond to a
positive fraction of the total sample size, the panel setup allows one to
consider also the case where $T_{b}^{0}$ can take any value in $\left[
1,\,T-1\right] $. The latter case can be studied under the asymptotics
framework with $N\rightarrow \infty $ and $T$ fixed or $N,\,T\rightarrow
\infty $ such that $T/N\rightarrow 0$. Importantly, only consistency can be
established under the latter scenario. For the derivation of the limiting
distribution one needs the standard assumption $T_{b}^{0}=T\lambda _{0}$.
Turning to the assumptions on the errors, Bai (2010) requires stationarity
of $\left\{ u_{it}\right\} $ in the time dimension and independence over $i$%
. It is argued that it can be relaxed without affecting the consistency
result, though for the asymptotic distribution one requires the
cross-sectional dependence to be not too strong. The assumption on the break
sizes is $\lim_{N\rightarrow \infty }N^{-1/2}\sum_{i=1}^{N}\left( \mu
_{i2}-\mu _{i1}\right) ^{2}=\infty $. To understand, note that if $\mu
_{i2}-\mu _{i1}$ were $i.i.d.$ random variables with positive variance then
the above limit with $N^{-1}$ replacing $N^{-1/2}$ should converge to a
positive constant. Thus, the condition does not require every unit to have a
break. The estimation method involves least-squares. For a given $1\leq
T_{b}\leq T-1$, define $\bar{y}_{i1}=T_{b}^{-1}\sum_{t=1}^{T_{b}}y_{it}$ and 
$\bar{y}_{i2}=\left( T-T_{b}\right) ^{-1}\sum_{t=T_{b}+1}^{T}y_{it}$ which
are the estimates of $\mu _{i1}$ and $\mu _{i2}$, respectively. Define the
sum of squared residuals for the $i^{th}$ equation as $S_{iT}\left(
T_{b}\right) =\sum_{t=1}^{T_{b}}\left( y_{it}-\bar{y}_{i1}\right)
^{2}+\sum_{t=T_{b}+1}^{T}\left( y_{it}-\bar{y}_{i2}\right) ^{2}$ for $%
T_{b}=1,\ldots ,T-1$ and $S_{iT}\left( T_{b}\right) =\sum_{t=1}^{T}\left(
y_{it}-\bar{y}_{i}\right) ^{2}$ for $T_{b}=T$ where $\bar{y}_{i}$ is the
whole sample mean. The least-squares estimator of the break point is defined
as $\hat{T}_{b}=\mathrm{argmin}_{1\leq T_{b}\leq T-1}SSR\left( T_{b}\right) $%
, where $SSR\left( T_{b}\right) =\sum_{i=1}^{N}S_{iT}\left( T_{b}\right) $.
The availability of panel data leads to a stronger result about the rate of
convergence. In the univariate case, we have $\hat{T}_{b}=T_{b}^{0}+O_{p}%
\left( 1\right) $. For panel data, under either fixed $T$ or $T\rightarrow
\infty $, $\hat{T}_{b}=T_{b}^{0}+o_{p}\left( 1\right) $ and results show
that with $N\rightarrow \infty $ the common break point can be estimated
precisely even with a regime having a single observation. The limiting
distribution is derived under a small shifts assumption with $\mu _{i2}-\mu
_{i1}=N^{-1/2}\Delta _{i}$, $\Delta _{i}>0$. Construction of the confidence
intervals requires simulations of the derived limiting distribution. Unlike
in the univariate case, a change in variable allowing to express it as a
function of quantities that can be consistently estimated cannot be
immediately carried out. Nonetheless, noting that a Gaussian random walk and
a standard Wiener process have the same distribution at integer times, one
can apply the change in variable argument leading to the same inference
procedure as in the univariate case, which now only holds approximately, so
that inference works as in the univariate setting. Finally, Bai (2010)
further considers the case of (possibly) simultaneous break in mean and
variance and proposes a QML method.

Kim (2011) studies least-squares estimation of a common deterministic time
trend break in large panels with a break either in the intercept, slope or
both with a general dependence structure in the errors. He models
cross-sectional dependence with a common factor structure and allows the
errors to be serially correlated in each equation: $u_{it}=\gamma
_{i}^{\prime }F_{t}+e_{ti}$ where $F_{t}$ is a vector of latent common
factors, $\gamma _{i}$ is a factor loading and $e_{it}$ is a unit specific
error. Under joint asymptotics $\left( T,\,N\right) \rightarrow \infty $,
serial correlation and cross-sectional dependence are shown to affect the
rate of convergence and the limiting distribution of the break point
estimator. As in Bai (2010), $\hat{T}_{b}$ can be consistent for $T_{b}^{0}$
even when $T$ is fixed, though it only holds when the $u_{it}$ are
independent across both $i$ and $t$. When the $e_{it}$ are serially
correlated and there are no common factors, the rate of convergence depends
on $N$ and is faster than in the univariate case. With common factors
generating strong cross sectional dependence, the rate of convergence does
not depend on $N$ and reduces to the univariate case (cf. Perron and Zhu,
2005). For fixed shifts, the limiting distribution depends on many elements;
e.g., the form of the break, the presence of common factors, stationary
versus integrated errors. For a joint broken trend, it can be normal. To
obtain a limit theory not depending on the exact distribution of the errors,
the break magnitudes need to converge to zero at a rate $N^{-1/2}$, in which
case asymptotically valid confidence intervals can computed via simulations.

Results pertaining to regression models using stationary panels were
obtained by Baltagi, Feng and Kao (2016). They consider large heterogeneous
panels with common correlated effects (CCE) and allowed for unknown common
structural breaks in the slopes. The CCE setting takes the following form: $%
y_{it}=x_{it}^{\prime }\beta (T_{b}^{0})+u_{it}$, where $u_{it}=\gamma
_{i}^{\prime }F_{t}+e_{it}$, $x_{it}=\Gamma _{i}^{\prime }F_{t}+v_{it}$, $%
e_{it},\,v_{it}$ are idiosyncratic errors and some or all components of $%
\beta (T_{b}^{0})$ differ pre and post-break. Due to the correlation between 
$x{}_{it}$ and $u{}_{it}$, least-squares for each cross-sectional regression
could be inconsistent. They use a least-squares method using augmented data
and confirm the result in Bai (2010) that the break point $T_{b}^{0}$ can be
consistently estimated as both $N$ and $T$ go to infinity. Common breaks in
panels were also considered by Qian and Su (2016) and Li, Qian and Su (2016)
who study estimation and inference with and without interactive fixed
effects using Lasso methods. Kim (2014) generalizes Kim (2011) by allowing a
factor structure in the error terms. Baltagi, Feng and Kao (2016) study
structural breaks in a heterogeneous large panel with interactive fixed
effects. They show the consistency of the estimated break fraction and break
date under some conditions.

Most of the work on structural breaks in panels focused on common breaks, in
which case $T_{b}^{0}$ itself can be consistently estimated and not only $%
\lambda _{b}^{0}$. One may infer that simply adding a cross-sectional
dimension yields more information and precise estimates. This is misleading
because the result crucially relies on the assumption that the break is
common to all units. Although this may be relevant in practice, the results
should be interpreted carefully.

\noindent \textbf{Continuous record asymptotics}. Casini and Perron (2017a)
consider an asymptotic framework based on a continuous-time approximation,
i.e., $T$ observations with a sampling interval $h$ over a fixed time span $%
\left[ 0,\,N\right] $, where $N=Th$, with $T\rightarrow \infty $ and $%
h\rightarrow 0$ with $N$ fixed. Liang et a. (2016) consider a similar,
though different, framework for the simple case of a change in mean and they
do not provide feasible versions for the construction of the confidence
sets, hence we follow the general approach of Casini and Perron (2017a) who
consider the following partial structural change model with a single break
point: 
\begin{equation}
Y_{t}=D_{t}^{\prime }\pi ^{0}+Z_{t}^{\prime }\delta _{1}^{0}+e_{t},\hspace{%
0.05in}(t\leq T_{b}^{0})\text{; \hspace{0.05in}}Y_{t}=D_{t}^{\prime }\pi
^{0}+Z_{t}^{\prime }\delta _{2}^{0}+e_{t},\hspace{0.05in}(t>T_{b}^{0}).
\label{Original SC Model}
\end{equation}%
where $\{D_{s},Z_{s},e_{s}\}_{s\geq 0}$ are continuous-time processes and we
observe realizations at discrete points of time, namely $\left\{
Y_{kh},D_{kh},Z_{kh};\,k=0,\ldots ,\,T=N/h\right\} $. For any process $X$,
we denote its ``increments'' by $\Delta _{h}X_{k}=X_{kh}-X_{\left(
k-1\right) h}$. For $k=1,\ldots ,T$, let $\Delta _{h}D_{k}=\mu
_{D,k}h+\Delta _{h}M_{D,k}$ and $\Delta _{h}Z_{k}=\mu _{Z,k}h+\Delta
_{h}M_{Z,k}$, where $\mu _{D,t},\,\mu _{Z,t}$ are the ``drifts'' and $%
M_{D,k},\,M_{Z,k}$ are continuous local martingales. They consider the
least-squares estimator of the break point and the analysis is valid for
general time series regression models including errors with correlation
and/or heteroskedasticity and lagged dependent variables. Under fixed
shifts, $\hat{T}_{b}-T_{b}^{0}=O_{p}\left( 1\right) $. Besides the usual
small shifts assumption, the limiting distribution is derived under the
additional assumption of increasing local variances around the true break
date. The continuous record asymptotic distribution of the least-squares
estimator is then given by 
\begin{equation}
Th(\hat{\lambda}_{b}-\lambda _{0})\Rightarrow \underset{v\in \lbrack
-N_{b}^{0}/(||\delta ^{0}||^{-2}\bar{\sigma}^{2}),\,\left(
N-N_{b}^{0}\right) /(||\delta ^{0}||^{-2}\bar{\sigma}^{2})]}{\mathrm{argmax}}%
\{-\left( \delta ^{0}\right) ^{\prime }\left\langle Z_{\Delta },\,Z_{\Delta
}\right\rangle \left( v\right) \delta ^{0}+2\left( \delta ^{0}\right)
^{\prime }W\left( v\right) \},  \label{CR-lim}
\end{equation}%
where $\left\langle Z_{\Delta },\,Z_{\Delta }\right\rangle \left( v\right) $
is the predictable quadratic variation process of $Z_{t}$, $W\left( v\right) 
$ is a two-sided centered Gaussian process and $\bar{\sigma}^{2}$ is the
limit of an estimate of the error innovation variance over $\left[ 0,\,N%
\right] $. The results (\ref{CR-lim}) is defined on a new ``fast time
scale''. The latter provides a better approximation to the properties of the
finite-sample distribution when $h$ is actually fixed. It is shown that the
continuous record asymptotic distribution provides a much better
approximation to the finite-sample distribution of the least-squares
estimator. The former is highly non-standard and captures the main
properties of the latter such as tri-modality, asymmetry and skewness. Thus,
basing inference on the continuous record asymptotic theory results in
inference procedures about the break date that perform better than existing
methods. As shown in Elliott and M\"{u}ller (2007) and Chang and Perron
(2017a), Bai's (1997) method for constructing confidence intervals for the
break date displays a coverage probability far from the nominal level when
the magnitude of the break is small. Casini and Perron (2017a) propose
constructing confidence sets by computing the Highest Density Regions (HDR)
of the density of the continuous record limiting distribution. Their
confidence sets are shown to provide accurate converge rates and relatively
short length of the confidence sets across different break magnitudes and
break locations when compared with those of Bai (1997), Elliot and M\"{u}%
ller (2007) and Eo and Morley (2015). In addition, Casini and Perron (2017b)
investigate a GL estimation and inference method, which involves
transforming the least-squares objective function into a proper distribution
(i.e., a Quasi-posterior) and minimizing the expected risk under a given
loss function. The analysis is carried out under continuous record
asymptotics. This yields a new estimator of the break shown to be more
accurate than the original least-squares estimator. The proposed confidence
sets, which use the HDR concept, also have coverage rates close to the
nominal level and of relatively small length whether the break is small or
large. The GL estimator based on the least-squares estimate was also
considered under large-$T$ asymptotics by Casini and Perron (2017c) who also
relate it to the distribution theory of Bayesian change-point estimators.

\noindent \textbf{Forecasting}. We first discuss the concept of forecast
failure (or breakdown) and describe methods proposed to detect changes in
the forecasting performance over time. Second, we discuss techniques to
compare the relative predictive ability of two competing forecast models in
an unstable environment. It is useful to clarify the purpose of forecast
breakdown tests. The aim is to assess retrospectively whether a given
forecasting model provides forecasts which show evidence of changes
(improvements or deterioration) with respect to some loss function. Since
the losses can change because of changes in the variance of the shocks
(e.g., good luck), detection of a forecast failure does not necessarily mean
that a forecast model should be abandoned. Care must be exercised to assess
the source of the changes. But if a model is shown to provide stable
forecasts, it can more safely be applied in real time. In practice, such
forecasts are made at the time of the last available data, using a fixed,
recursive or rolling window. Hence, there is a natural separation between
the in-sample and out-of-sample periods simply dictated by the last data
point. Such is not the case when trying to assess retrospectively whether a
given model provides stable forecasts. There is then the need for a somewhat
artificial separation between the in and out-of-sample periods at some date
labelled $T_{m}$, say. This separation date should be such that the model in
the in-sample period is stable in some sense, e.g., yielding stable
forecasts. This can, however, create problems; e.g., one needs a truncation
point $T_{m}$ to assess forecast failures but the choice of this value is
itself predicated on some knowledge of stability.

The forecast failure test of Giacomini and Rossi (2009), GR (2009)
hereafter, is a global and retrospective test which compares the in-sample
average with the out-of-sample average of the sequence of forecast losses.
Adopting the same notation as in the previous section, we have $k=1,\ldots
,\,T$ observations with a sampling frequency $h$ over the time span $\left[
0,\,N\right] $ with $N=Th$. We recover the setting of GR (2009) by setting $%
h=1$ in what follows. Define at time $\left( k+\tau \right) h$ a surprise
loss given by the deviation between the time-$\left( k+\tau \right) h$
out-of-sample loss and the average in-sample loss: $SL_{\left( k+\tau
\right) h}(\hat{\beta}_{k})=L_{\left( k+\tau \right) h}(\hat{\beta}_{k})-%
\bar{L}_{kh}(\hat{\beta}_{k}),$ for $k=T_{m},\ldots ,\,T-\tau $, where $\bar{%
L}_{kh}(\hat{\beta}_{k})$ is the average in-sample loss computed according
to the specific forecasting scheme, where $\hat{\beta}_{k}$ is some
estimator of the model parameters and $T_{m}$ is the in-sample size. One can
then define the average of the out-of-sample surprise losses $\overline{SL}{}%
_{N_{0}}(\hat{\beta}_{k})=N_{0}^{-1}\sum_{k=T_{m}}^{T-\tau }SL_{\left(
k+\tau \right) h}(\hat{\beta}_{k}),$ where $N_{0}=N-N_{in}-h$ denotes the
time span of the out-of-sample window and $N_{\mathrm{in}}=T_{m}h$. GR
(2009) observed that under the hypothesis of no forecast instability $%
\overline{SL}{}_{N_{0}}$ should have zero mean (i.e., no systematic surprise
losses in the out-of-sample window). Under the null hypothesis of no
forecast failure, the GR (2009) test $t_{GR}=N_{0}^{1/2}\overline{SL}{}%
_{N_{0}}(\hat{\beta}_{k})/\hat{\sigma}_{T_{m},T_{n}}$ follows asymptotically
a standard normal distribution.

Casini (2017) extends the analysis by considering a continuous-time
asymptotic framework and partitioning the out-of-sample into $%
m_{T}=\left\lfloor T_{n}/n_{T}\right\rfloor $ blocks each containing $n_{T}$
observations. Let $B_{h,b}=n_{T}^{-1}\tsum\nolimits_{j=1}^{n_{T}}SL_{T_{m}+%
\tau +bn_{T}+j-1}(\hat{\beta})$ and $\bar{B}_{h,b}=n_{T}^{-1}\tsum%
\nolimits_{j=1}^{n_{T}}L_{\left( T_{m}+\tau +bn_{T}+j-1\right) h}(\hat{\beta}%
)$ for $b=0,\ldots ,\,\left\lfloor T_{n}/n_{T}\right\rfloor -1$ with $T_{n}$
the out-of-sample size. The test statistic is, $Q_{max,h}\left( T_{n},\,\tau
\right) =\nu _{L}^{-1}\max_{b=0,\ldots ,\,\left\lfloor
T_{n}/n_{T}\right\rfloor -2}\left| B_{h,b+1}-B_{h,b}\right| $ where $\nu
_{L} $ is the square root of the asymptotic variance of the test. The test
partitions the out-of-sample window into $m_{T}$ blocks of asymptotically
vanishing length $\left[ bn_{T}h,\,\left( b+1\right) n_{T}h\right] $ and $%
B_{h,b}$ is a local average of the surprise losses within the block $b$. The
test $Q_{max,h}\left( T_{n},\,\tau \right) $ takes on a large value if there
is a large deviation $B_{h,b+1}-B_{h,b}$, which suggests a discontinuity or
non-smooth shift in the surprise losses close to time $bn_{T}h$ and thus it
provides evidence against the null. Simulations show that the test of GR
(2009) and Casini (2017) both have good power properties when the
instability is long-lasting while the latter performs better when the
instability is short-lived.

Perron and Yamamoto (2017) adapt the classical structural change tests to
the forecast failure context. First, they recommend that all tests should be
carried with a fixed scheme to have best power, which ensures the maximum
difference between the fitted in and out-of sample means of the losses.
There are contamination issues under the rolling and recursive scheme that
induce power losses. With such a fixed scheme, GR's (2009) test is simply a
Wald test for a one-time change in the mean of the total (the in-sample plus
out-of-sample) losses at a known break date $T_{m}$. To alleviate this
problem, which leads to important losses in power when the break in
forecasting performance is not exactly at $T_{m}$, one can follow Inoue and
Rossi (2012) and consider maximizing the GR (2009)\ test over all possible
values of $T_{m}$ within a pre-specified range. This then corresponds to a
sup-Wald test for a single change at some date constrained to be the
separation point between the in and out-of-sample periods. The test is still
not immune to non-monotonic power problems when multiple changes occur.
Hence, Perron and Yamamoto (2017) propose a Double sup-Wald test which for
each $T_{m}\in \left[ T_{0},\,T_{1}\right] $ performs a sup-Wald test for a
change in the mean of the out-of-sample losses and takes the maximum of such
tests over the range $T_{m}\in \left[ T_{0},\,T_{1}\right] $: $%
DSW=\max_{T_{m}\in \left[ T_{0},\,T_{1}\right] }SW_{L^{o}\left( T_{m}\right)
}$, where $SW_{L^{o}\left( T_{m}\right) }$ is the sup-Wald test for a change
in the mean of the out-of-sample loss series $L_{t}^{o}(\hat{\beta})$ for $%
t=T_{m}+\tau ,\ldots ,\,T$, defined by 
\begin{equation*}
SW_{L^{o}\left( m\right) }=\max_{T_{b}\left( T_{m}\right) \in \left[
T_{m}+\epsilon T,\,T_{m}+\left( 1-\epsilon \right) T\right]
}[SSR_{L^{o}\left( m\right) }-SSR\left( T_{b}\left( T_{m}\right) \right)
_{L^{o}\left( T_{m}\right) }]/\hat{V}_{L^{o}\left( T_{m}\right) },
\end{equation*}%
where $SSR_{L^{o}\left( T_{m}\right) }$ is the unrestricted sum of squares, $%
SSR\left( T_{b}\left( T_{m}\right) \right) _{L^{o}\left( T_{m}\right) }$ is
the sum of squared residuals assuming a one-time change at time $T_{b}\left(
T_{m}\right) $, and $\hat{V}_{L^{o}\left( T_{m}\right) }$ is the long-run
variance estimate of the out-of-sample loss series. In addition, Perron and
Yamamoto (2017) propose to work directly with the total loss series $L\left(
T_{m}\right) $ to define the Total Loss Sup-Wald test (TLSW) and the Total
Loss UDmax test (TLUD). Using extensive simulations, based on the original
design of GR (2009) which involves single and multiple changes in the
regression parameters and/or the variance of the errors, they show that with
forecasting models potentially involving lagged dependent variables, the
only tests having a monotonic power function for all DGPs are the Double
sup-Wald and Total Loss UDmax tests, constructed with a fixed forecasting
window scheme (Casini's (2017) test was not included in the simulations).

Next, we turn to testing for forecast comparisons in unstable environment.
Here, the goal is to determine the relative out-of-sample predictive ability
between two competing models in the presence of possible breaks. Giacomini
and Rossi (2010) propose two tests: the Fluctuation test and the One-time
Reversal test. The former tests whether the local relative forecasting
performance equals zero at each point in time whereas the One-time Reversal
tests the null hypothesis that the two models perform equally well at each
point in time against the alternative that there is a break in the relative
performance. Here we discuss the Fluctuation test only. Suppose we compare
two $\tau $-step ahead forecast models for the scalar $y_{k}$. The first
model is characterized by a parameter $\theta $ and the second model by a
parameter $\gamma $. The relative performance is evaluated by a sequence of
out-of-sample loss differences $\{\Delta L_{k}(\hat{\theta}_{k-\tau ,T_{m}},%
\hat{\gamma}_{k-\tau ,T_{m}})\}_{k=T_{m}+\tau }^{T}$, where $\Delta L_{k}(%
\hat{\theta}_{k-\tau ,T_{m}},\hat{\gamma}_{k-\tau ,T_{m}})=L^{\left(
1\right) }(y_{k},\hat{\theta}_{k-\tau ,T_{m}})-L^{\left( 2\right) }(y_{k},%
\hat{\gamma}_{k-\tau ,T_{m}})$. The expressions for the estimators $\hat{%
\theta}_{k-\tau ,T_{m}}$ and $\hat{\gamma}_{k-\tau ,T_{m}}$ depend on the
forecasting scheme. The Fluctuation test is Fluct$_{k,T_{m}}^{o}=m^{-1}\hat{%
\sigma}^{-1}\tsum\nolimits_{k=t-m/2}^{k+m/2-1}\Delta L_{k}(\hat{\theta}%
_{k-\tau ,m},\,\hat{\gamma}_{k-\tau ,m})$, for $k=T_{n}+\tau +m/2,\ldots
,\,T-m/2+1$, where $\hat{\sigma}^{2}$ is an estimate of the long-run
variance of the sequence of out-of-sample losses and $m$ is the size of the
window. The asymptotic null distribution is non-standard and critical values
are computed by simulations. The test has good finite-sample properties
under serially uncorrelated losses when scaled by an estimate of the
variance instead of the long-run variance. However, Martins and Perron
(2016) show that the test suffers from non-monotonic power when constructed
with the long-run variance estimate, as it should be, whether or not the
sequence of loss differences exhibit serial correlation. They propose using
simple structural change tests such as the sup-Wald test of Andrews (1993)
and the UDmax test of Bai and Perron (1998). These are preferred since they
have the highest monotonic power even with a long-run variance constructed
with a constrained small bandwidth. Finally, Fossati (2017) notices that
when the predictive ability is state-dependent (e.g., recessions versus
expansions), then taking account of such property by using a test based on
Markov Regime-Switching can be a useful alternative.

Additional work and surveys that relate to the issue of testing for
structural changes in forecasting and/or forecasting allowing for possible
in and out-of-sample changes include, among others, Clements and Hendry
(1998a,b, 2006), Pesaran, Pettenuzzo, and Timmermann (2006), Banerjee,
Marcellino, and Masten (2008), Rossi (2013a), Giacomini (2015), Giacomini
and Rossi (2015) and Xu and Perron (2017).

\end{document}